# Orientational Ordering of Passivating Ligands on CdS Nanorods in Solution Generates Strong Rod-Rod Interactions

*Asaph Widmer-Cooper*[*,1,3] *and Phillip Geissler*[2,3,4]

[1]School of Chemistry, University of Sydney, Sydney, NSW 2006, Australia, [2] Department of Chemistry, University of California Berkeley, Berkeley, California 94720, and [3]Materials Sciences Division and [4]Chemical Sciences Division, Lawrence Berkeley National Laboratory, Berkeley, California 94720

Corresponding author: asaph.widmer-cooper@sydney.edu.au.

**ABSTRACT**

We present the first nearly atomistic molecular dynamics study of nanorod-nanorod association in explicit solvent, showing that inter-rod forces can be dominated by microscopic factors absent in common continuum descriptions. Specifically, we find that alkane ligands on faceted CdS nanorods in *n*-hexane undergo a temperature-dependent order-disorder transition akin to that of self-assembled monolayers on macroscopic substrates. This collective ligand alignment organizes nearby solvent molecules, strongly influencing the statistics of rod-rod separation. The strong temperature-dependence of this mechanism could be exploited in the laboratory to manipulate and optimize the assembly of ordered structures.

Colloidal rods can now be made from a wide range of materials including viruses, metals, semiconductors, and carbon nanotubes,[1–4] and have been assembled into many different types of structures, including strings, spherical shells, surface smectics, and lamellar tactoids.[5–12] The potential technological utility of such assemblies (e.g., in making printable nanostructured solar cells, photoelectrochemical devices, and diodes[13–17]) has inspired a significant effort to understand and control their formation. As one example, Refs. [15,17,18], describe recent successes in covering large surfaces with assemblies of perpendicularly aligned rods.

This paper concerns one key factor determining when and how nanorods assemble, namely, the nature and strength of interactions between them. Previous work has highlighted the sensitivity of assembly dynamics to the strength and specificity of forces among nanoparticles.[19] In particular, attractive interactions between colloidal rods influence not only their phase behavior[5,20,21] but also the types of non-equilibrium assemblies they can form. One way to induce attractions controllably is by adding molecules that act as depleting agents. Such depletion attraction can lead to the formation of long-lived lamellar tactoids,[5,21,22] consisting of a monolayer of rods aligned parallel to one another and perpendicular to the plane of the lamellae. Cases in which other forces between colloidal rods have been directly measured or calculated in detail, however, are few. It has therefore remained unclear how to best engineer and tune their interactions, which may involve a complex interplay between direct rod-rod forces and those mediated by fluctuations in their environment.

As a specific example with practical interest, we focus here on nanocrystals whose exteriors are covered with surfactant-like ligands. Several theoretical approaches have been taken to estimate interactions between such particles, idealizing molecular structure and fluctuations in different ways. One approach is to ignore ligand and solvent fluctuations completely, and model the ligands and solvent as continuous dielectric media.[23] This assumes dominance of the vdW interaction between the crystalline cores and of any additional depletion-attraction. A more sophisticated approach is to treat the ligands as a continuous dielectric shell around the crystalline core and account for the osmotic repulsion between the ligand shells in an approximate analytical way.[24] This level of detail reveals that even when the ligand-ligand attraction is weak it can make a significant contribution to the interparticle force. One could take this continuum approach one step further, allowing the ligands[25] and solvent[26] to have non-uniform density profiles, and thus account for the soft nature of the ligand-solvent interface and solvent structuring. None of these methods, however, addresses contributions from the passivating ligands more complicated than dispersion attraction or steric repulsion that would arise from a continuous medium. This neglect of complexity is motivated in part by an assumption that the passivating layer lacks significant structural organization – it is viewed essentially as a dense but disordered collection of short chain molecules.

Spectroscopic measurements on small Au particles coated with alkylthiol molecules suggest that these ligands are indeed disordered in nonpolar solvent.[27] Molecular simulations are consistent with this observation[28] and, in addition, predict that the interaction between small Au dots covered by disordered ligands is purely repulsive in both polar[29] and nonpolar[30] solvents. However, there is also evidence that ligands on nanoparticles can align into ordered patches in poor solvents[28,31] and on dry particles[32–36], and that such ordering can affect the relative stability of different nanoparticle assemblies[32,37]. These results suggest that in certain situations nanoparticles may have much in common with macroscopic self-assembled monolayers (SAMs)[38–40], whose interactions can be richly influenced by ligand organization[41].

Here we report on detailed computer simulations of nanorods in solution, which point to a role for ligands in mediating rod-rod interactions that is both more subtle and more potent than has previously been described. Specifically, we use large-scale molecular dynamics (MD) simulations to study the temperature-dependent structure of octadecyl ligands and of *n*-hexane solvent around 4x20 nm CdS nanocrystals, as well as the effect they have on the rod-rod interaction. We find that surfactant ligands on nanorods can undergo a temperature-dependent order-disorder transition in solution. This ordering phenomenon effects considerable changes in the solvent density near the particle surface, much as with a classic self-assembled monolayer (SAM)[38,39]. Changes in the ligand and solvent structure can in turn change the force acting between a pair of rods from purely repulsive to strongly attractive, in sharp contrast to previous results for Au particles with small facets. These subtle but significant changes in the free energy of rod-rod association should have a strong impact on how CdS nanorods, and other particles with extended facets, assemble from solution.

Our microscopic model of a passivated CdS nanorod, motivated by available structural information, is illustrated in Figure 1. In detail, the CdS core was modeled as a static wurzite lattice, in the shape of a 4x20 nm hexagonal prism with six (100) facets on the sides, a Cd-terminated (001) facet on one end and a S-terminated (001) facet on the other. These are the dominant facets observed in HRTEM images of CdS and CdSe nanorods[42]. We dressed our model nanocrystals with octadecyl ligands at high surface coverage as suggested by experiment and electronic structure calculations (see Supporting Information): with one ligand bound to: (a) each surface Cd atom on the (100) side facets, (b) 3 out of every group of 4 Cd atoms on the Cd-terminated (001) facet, and (c) every second S atom on the S-terminated (001) facet (see Figure 1). Our model does not explicitly include the phosphonate headgroups of the ligands, which are deeply buried within a dense shell of alkyl chains. Aside from anchoring chains to the corresponding Cd atoms, these hardly exposed headgroups should only weakly impact rod-rod association. In our model, the Cd-$CH_2$ bond is treated identically to the $CH_x$-$CH_x$ bonds (see below).

The ligands and the hexane solvent were modeled using a united-atom potential, which represents each $CH_x$ group with a single particle. Interactions between these coarse-grained particles include volume exclusion and dispersion as described by the Lennard-Jones (LJ) potential and, within each molecule, bond stretching, bond bending, and dihedral torsion terms. This description extends slightly the TraPPE potential developed by Martin and Siepmann for studying the phase behavior of alkanes[43], adding a bond stretching term (taken from ref. [44]) for compatibility with the MD package LAMMPS[45]. The model CdS-$CH_x$ interaction was adapted from a LJ potential previously developed for Au nanoparticles[46]. Specifically, LJ parameters (Table S1) were modified to reflect the Hamaker constant and density of CdS[47] rather than Au. The crystalline CdS cores of different nanorods interact in our model via the Hamaker potential[48], with a Hamaker constant of 0.7125 eV [47]. In practice, we find that the core-core interaction makes only a negligible contribution to the total rod-rod interaction at accessible separations for the 4x20 nm nanocrystals considered here. Further details of our model and simulation methods are given in the SI.

Molecular dynamics (MD) simulations on systems of up to 350,000 particles were performed using LAMMPS[45], for a variety of temperatures and inter-rod distances. In all cases we fixed the position and orientation of each rod, as well as the total number of solvent and ligand molecules. Configurations with appropriate density were constructed by allowing the ligands and initially dilute solvent to relax while slowly compressing the periodic simulation cell. More specifically, compression ceased when regions of space distant from rod surfaces contained an average number of hexane molecules per unit volume matching the experimental density of pure *n*-hexane at the relevant temperature. Systems were then equilibrated with fixed volume and constant temperature, maintained with a Nosé-Hoover thermostat, for at least 1 ns. Subsequent production runs at fixed temperature and volume were at least 500 ps (and up to 10 ns) in duration for each set of thermodynamic conditions and rod geometries.

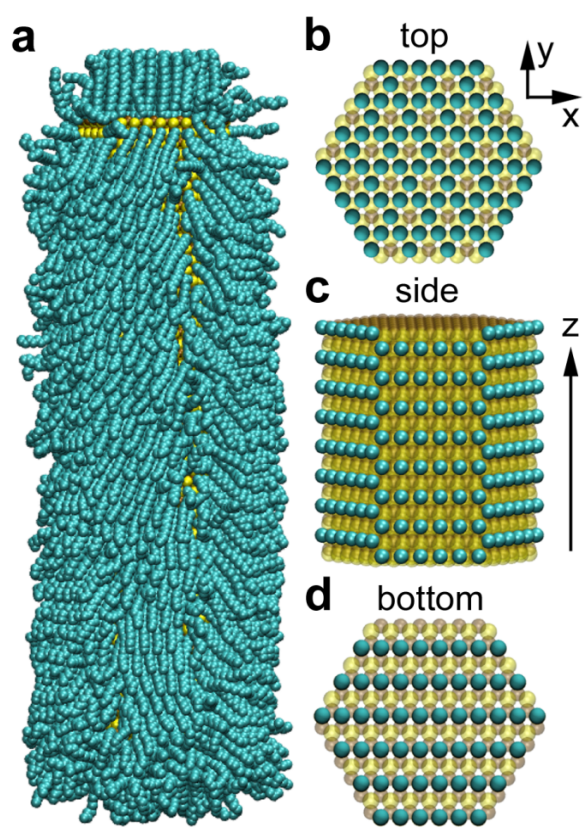

**Figure 1.** (a) A 4 x 20 nm CdS nanorod passivated with octadecyl ligands and equilibrated in hexane at 300 K (solvent not shown). Views of (b) the top (001)Cd facet, (c) the side (100) facets, and (d) the bottom (001)S facet, showing the different patterns of ligand coverage on these surfaces (Cd = brown, S = yellow, $CH_x$ = cyan). For clarity, only the terminal -$CH_2$- group is shown in (b)-(d).

We used constrained MD to calculate the potential of mean force (PMF) between two parallel rods, separated in the direction perpendicular to their long axis and with (100) facets facing one another. The mean force $F_{mean}$ between such nanorods held at a distance $r$ is given by the average force in the direction of their connecting line:

$$F_{mean}(r) = \frac{1}{2}\langle (F_2 - F_1) \cdot \hat{r} \rangle_{NVT} \quad (1)$$

where $F_1$ and $F_2$ are the total forces acting on the first and second nanorod, respectively, $\hat{r}$ is the unit vector pointing from one rod's center to the other's, and angular brackets denote an average in the canonical ensemble. The PMF is then given by:

$$\phi_{MF}(r) = \int_r^\infty F_{mean}(s)ds \quad (2)$$

Our simulations highlight similarities between the passivated facets of a nanorod and macroscopic surfaces coated with small chain molecules. Specifically, they suggest that side facets of a 4x20 nm nanorod are sufficiently large to allow substantial ordering of ligands, even in good solvents, in a manner highly reminiscent of macroscopic SAMs. Simulations of an extended CdS surface passivated by the same ligands in hexane exhibit strong ligand alignment at room temperature, which weakens sharply near 365 K.[49] As with SAMs on more conventional substrates, the highly aligned state is stabilized by favorable van der Waals attractions among ligands in all-trans conformations, which pack tightly. The passivation layer on our nanorods undergoes a highly cooperative transition of a similar nature, featuring pronounced changes in both the density of the ligands and their orientational statistics.

We describe the orientation of ligands on the side facets of the rods according to their tilt away from the normal vector of the surface to which they are bound. This tilt generally has a component along the long axis of the rod (vertical tilt angle $\theta_z$), and a component along the facet's short axis (horizontal tilt angle $\theta_x$), see Figure 3a. High- and low-temperature states of the nanorod's passivation layer differ distinctly in the statistics of both these tilt angles. On the time scale of our simulations, horizontal symmetry is in fact broken spontaneously on each side facet at low temperature. For this reason we refer to the low- and high-temperature states as ordered and disordered, respectively.

The vertical tilt of ligands on the nanorod is strong in the ordered state, $\langle \theta_z \rangle \approx 44$, as is evident from the configuration depicted in Figure 1a, which is representative of room temperature (300 K). This average orientation is nearly identical to the one we calculate for the nanorod's macroscopic counterpart under the same conditions, in which ligands are bound to a flat and periodically replicated (100) surface (the same facet as the sides of the nanorod). Alignment weakens in the disordered state but remains on average nonzero, $\langle \theta_z \rangle \approx 15°$, at 340 K. The average ligand density plotted in Figure 2, together with the temperature dependence of ligand orientation in Figure 3, provide a more quantitative view of this transition.

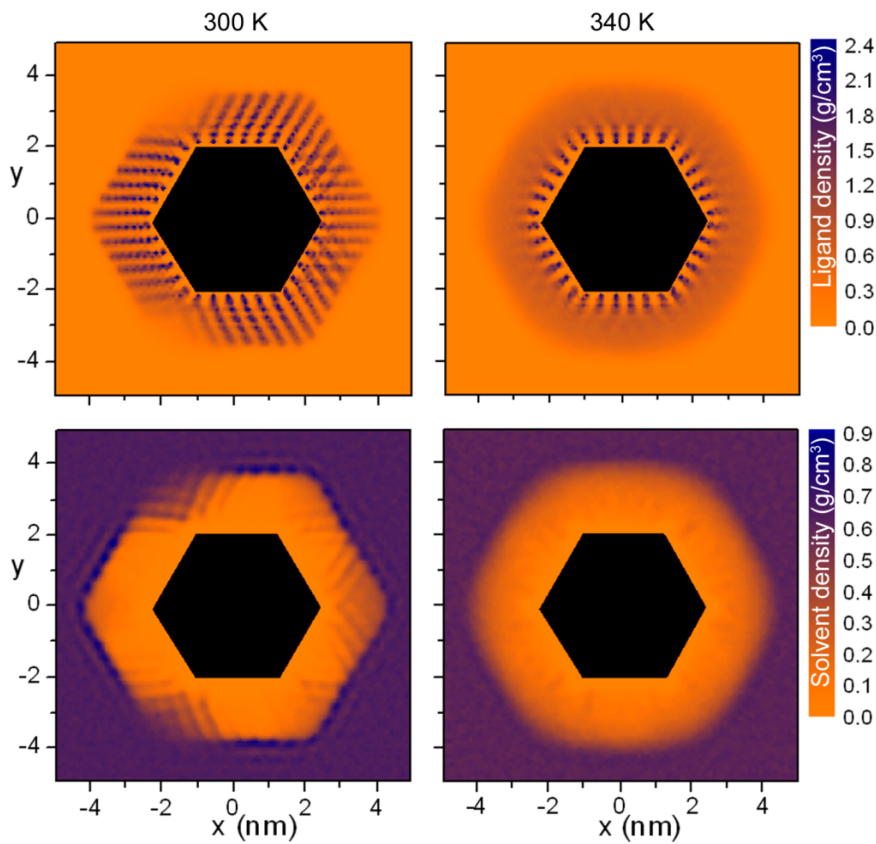

**Figure 2.** Plots of the ligand density (top) and solvent density (bottom), averaged over time and along the 20 nm length of the CdS core, for isolated rods in hexane at 300 K (left) and 340 K (right). As the ligands order, the ligand-solvent interface becomes more sharply defined and the solvent near the ordered ligand regions aligns into layers. The *x* and *y* scales are in nm and the contour scale is in g/cm$^3$. The black hexagon indicates the position of the CdS core. The actual simulation cell was larger than the region shown by an additional 1.5 nm in all directions.

While revealing thermodynamic behavior with close macroscopic analogies, our computational results also draw attention to significant finite size effects on the structure and stability of the ordered state. Loss of strong ligand alignment upon heating (see Figure 2b) begins at a lower temperature (~310 K) and proceeds more smoothly (over a range of ~20 K) than in the case of a flat CdS (100) surface[49] (where melting occurs sharply between 360 K and 365 K). We also find that the different degrees of ligand coverage on the top and bottom facets of our model nanorod break symmetry along the rod axis, with the free ligand ends always biased towards the more sparsely passivated S-terminal end (see Figure 1).

The finite dimension of the rod's side facets in the shorter x-direction (see Figure 3a) lead to more substantial differences compared to a macroscopic SAM. Ligands on each side facet tend to align with those on an adjacent side facet in the ordered state, so that $\langle \theta_x \rangle$ is nonzero, changing sign from one facet to the next (see Figures 2 and 3c). As a result, two aligned domains effectively merge into one, reducing their corresponding interfacial free energy. This alignment requires that the direction of ordering, relative to the facets' lattice vectors, differ from that of an extended SAM, for which the $\theta_x$ distribution remains a zero-centered Gaussian at all temperatures[49]. For a sufficiently broad rod, the advantage of merging adjacent domains (whose interfaces scale in extent with the length of a side facet) would be overwhelmed by the accumulated strain of rotating ligands away from their macroscopically preferred orientation (whose cost scales with the side facet area). On the 4x20 nm scale of our rods, however, merging domains is evidently favored strongly. Similar merging of domains has also been observed for ~3 nm icosahedral Au particles covered in dodecanethiol ligands in vacuum[34]. For the nanorod, this merging produces three distinctly aligned regions, divided by three furrows extending down the length of the rod. Overall, we find that the average potential energy $U_{LL}$ of pair interactions within the ligand shell decreases by ~2.1 kcal/mol per ligand across the transition (see Figure 3b), indicating a substantial concomitant gain of entropy of at least 3.5 $k_B$ per ligand upon disordering.

These changes in ligand ordering are accompanied by changes in the solvent structure in the vicinity of the nanorod. As the ligands order, the ligand-solvent interface becomes more sharply defined and the solvent near the ordered ligand regions aligns into layers, resulting in density oscillations, along the surface normal of the closest facet, up to 3 solvent layers deep (see Figure 2). The periodicity of these oscillations, ~0.5 nm, is consistent with alignment of hexane molecules roughly parallel to the rod surface. We have observed similar oscillations for an extended (100) surface.[49] Within the furrows separating well-ordered ligand domains on the nanorod's surface, more complicated solvent patterns result from interference between the densely packed regions on either side.

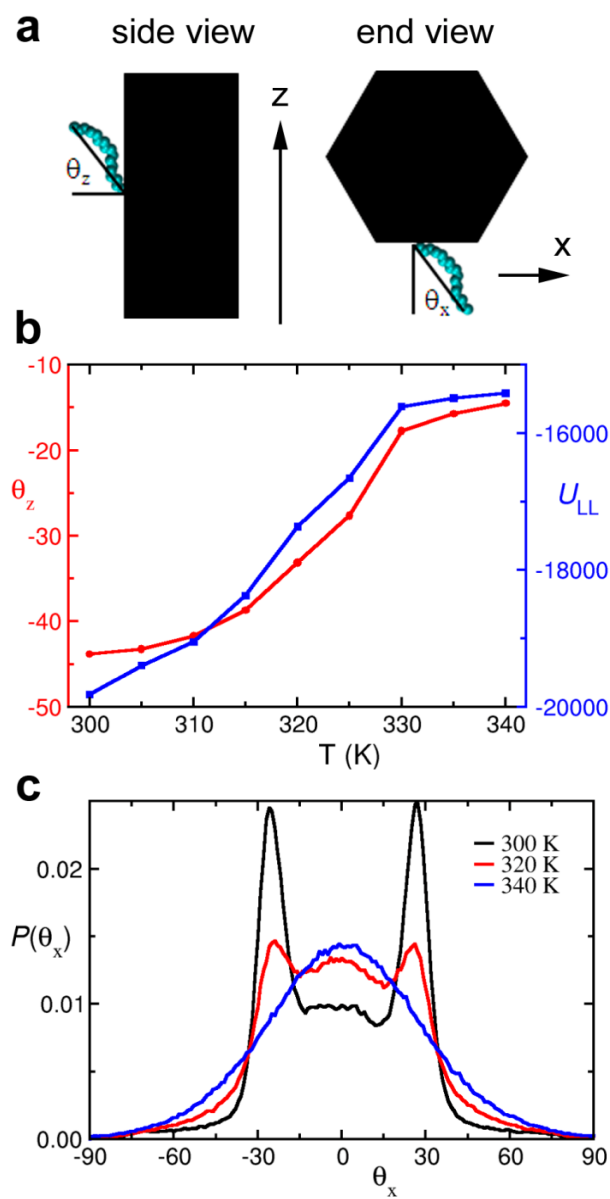

**Figure 3.** Temperature dependence of ligand ordering for an isolated rod in *n*-hexane. (a) Definition of ligand angles, (b) the mean polar angle $\langle \theta_z \rangle$ and ligand-ligand pair interaction energy $\langle U_{LL} \rangle$ (kcal/mol), and (c) the azimuthal angle distributions $P(\theta_x)$ at 300 K, 320 K and 340 K. The uncertainties in (b) are smaller than the symbol sizes.

The influence of ligand ordering on association between nanorods is profound and multifaceted. The effective interactions governing rod-rod separation include contributions from all fluctuating degrees of freedom that couple to rod translations. These numerous microscopic variables may in turn be biased by ligand structure, possibly in subtle ways. Most straightforwardly, a tightly packed ligand shell constitutes a more polarizable material than a disordered shell. Ligand ordering thus shifts the balance of van der Waals forces among ligands, solvent, and nanocrystal cores. Furthermore, a dense shell occupies less volume and therefore permits closer approach between two rods.

Additional effects of ligand structure on the mean force between two rods are principally entropic in character. Ligand chains in a disordered shell possess considerable conformational freedom. Interpenetration of two rods' passivation layers reduces the volume accessible to these fluctuations, decreasing entropy and thus inducing an average repulsion between rods. This kind of entropic repulsion[50] is largely absent for ordered shells, which possess little conformational entropy to begin with. More subtly, ligand structure influences the spatial arrangement of solvent molecules between two rods, which for an ordered shell can give rise to preferred separation distances consistent with the solvent's natural structure.

The potentials of mean force (PMFs) we have computed for the approach of two parallel rods exhibit signatures of these ligand-mediated interaction mechanisms. Figure 4a shows the PMF for rods in hexane solvent at three different temperatures spanning the transition from well-ordered to disordered passivation layers. At the highest temperature, 340 K, the mean force is purely repulsive (within statistical uncertainty of ~4 $k_\mathrm{B}T$/Å). The range of this repulsion, about 4 nm from one crystal surface to the other, corresponds to the combined extent of the two ligand shells. In other words, substantially repulsive forces emerge as soon as the passivation layers come into contact. The entropic nature of these forces is highlighted by the fact that average energies of ligand-ligand interactions *decrease* monotonically as the rods approach, i.e. the vdW attraction between the ligands becomes stronger. The strength of this entropic repulsion prohibits significant interpenetration of disordered shells – the work required to enforce an overlap of 0.2 nm is ~10 $k_\mathrm{B}T$. In *n*-hexane, the interaction between 1.8-2.7 nm icosahedral Au particles covered in disordered ligands has also been shown to be purely repulsive.[30]

At 300 K, where ligand ordering is essentially complete, changes in free energy as the two rods approach are dramatically different from the high-temperature case described above. Effective attractions between the rods at the lower temperature are due in part to their more compact ligand shells. This contraction of the passivating layers allows the rods to approach more closely before their ligands overlap, resulting in a stronger maximum attraction between the passivating layers than at high temperature (see Figure S1). More strikingly, strong oscillations appear in the lower-temperature PMF, with a deep local minimum at 3.3 nm separation and a substantial barrier at 3.6 nm. These features arise from spatial organization not only in the ligand shell but also in surrounding solvent. The periodicity of PMF oscillations (0.4-0.5 nm) corresponds to the spacing between solvent layers in the vicinity of an ordered ligand layer (see Figure 2c). Free energy minima occur at rod separations for which an integer number of solvent layers fit naturally between the rods (Figure 4b).

Similarly oscillating interactions mediated by solvent structuring have been calculated for unpassivated nanoparticles with solvophilic surfaces[51–53]. Also in close analogy, surface forces between macroscopic SAMs[41] in good solvents are known to be sensitive to ligand ordering. Specifically, strong attractive forces measured in the case of highly ordered ligands attenuate significantly when ligand alignment weakens. For solvents comprising small, simple molecules (e.g., *n*-alkanes and benzene), such measurements have revealed forces that oscillate with separation distance, just like those we have determined for ligand-coated nanorods.

At temperatures intermediate between 300 K and 340 K, proximity to the order-disorder transition creates a rich interplay between ligand shell structure and rod association. In this regime, ligand organization on an isolated rod is highly susceptible to external perturbations. Depending on the separation $r$, the presence of another rod nearby can therefore suffice to stabilize either more or less ordered states. At each separation, the resulting degree of ligand order in turn determines whether low-or high-temperature behavior dominates the PMF. Results for 320 K (marking the midpoint of the transition, as measured by the mean z-angle) are shown in Figure 4. As the rods approach from large $r$, ligand layers on their opposing facets initially extend toward one another and resemble the disordered high-temperature state (see Figures S2, S3 and S5); as at 340 K, the intervening solvent is unstructured and an entropic repulsive force develops over $r \approx 3.5 - 4$ nm. At closer

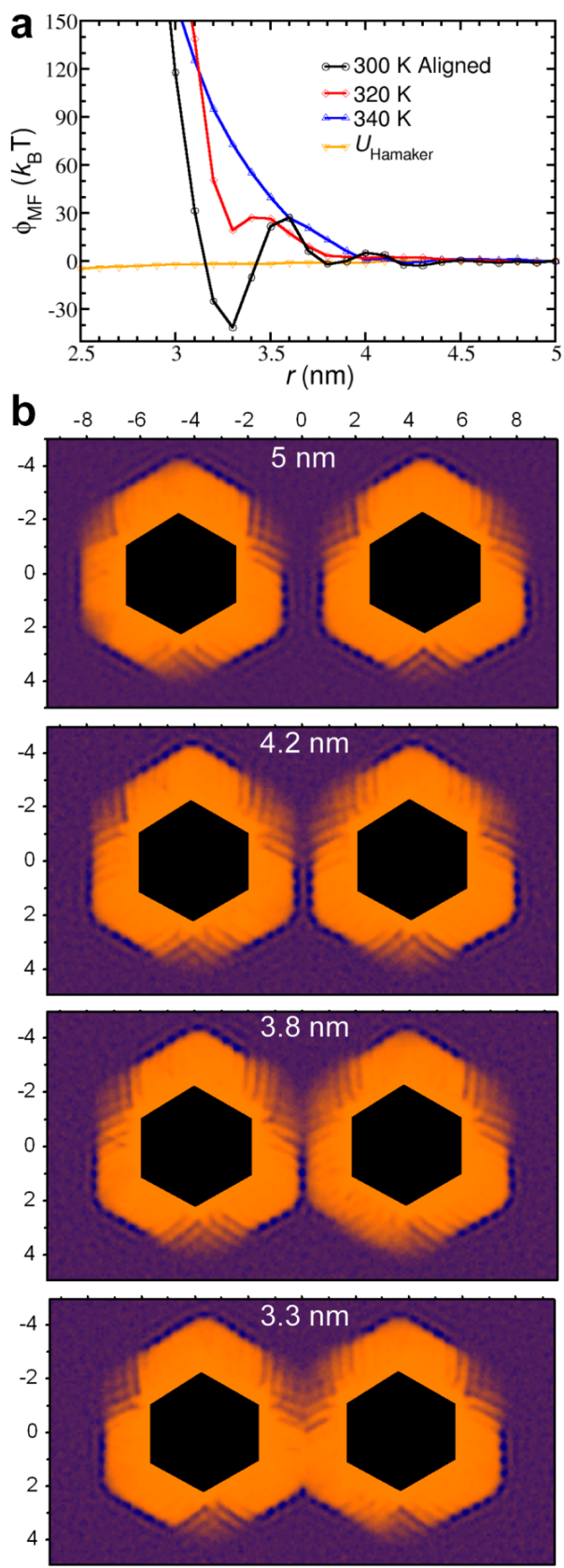

**Figure 4.** (a) Potentials of mean force $\phi_{MF}(r)$ for two parallel rods at 300 K, 320 K, and 340 K in *n*-hexane, where *r* is the distance between the opposing crystal facets. (b) Solvent densities at 300 K, averaged over time

and along the 20 nm length of the CdS core, for rod-rod separations corresponding to minima in $\phi_{MF}(r)$. The scales are the same as for the solvent density in Figure 2.

range these ligand layers become more strongly ordered, producing a minimum in the PMF at $r \approx 3.3$ nm corresponding to contact of the now-compact ligand shells. This metastable separation lies, however, well above the free energy of well-separated rods. The physically relevant portion of the PMF at 320 K, which is purely repulsive, more closely resembles the high-temperature case. The onset of strong repulsion, however, is shifted to closer separation, as might be expected from the initially denser ligand layers.

In all cases, the interaction energy between ligands on opposite rods $U_{Lig1-Lig2}$ becomes increasingly negative as the rods approach one another and the ligands interact more strongly. The ligand-ligand repulsion that sets in at close range must therefore be an entropic effect, most likely due to suppression of the ligand motion as the rods come into contact. In contrast to $U_{Lig1-Lig2}$, the ligand-solvent interaction energy $U_{Lig-Solv}$ is negative at all separations but becomes less negative as the rods approach, due to a reduction in the number of strong ligand-solvent interactions when solvent is excluded from the region between the rods. As a result, the attraction between the ligands and the solvent effectively pushes the rods apart, as reflected by a ligand-solvent force $F_{Lig-Solv}$ that is repulsive at most separations.

Ordering of the ligand shell at low temperature also breaks the six-fold rotational symmetry about the rod's long axis, yielding two distinct ways in which a pair of parallel rods can approach. In one case, ligand layers on the opposing faces are aligned with each other, i.e., their furrows of sparse ligand density are adjacent at contact (as are the regions where ligand termini are concentrated). The low-temperature PMF described above corresponds to this "aligned" arrangement, which is shown in Figure 4b. In the second case, opposing faces are misaligned, i.e., the ligand furrow on one face lines up not with the opposing furrow but instead with the opposing region of concentrated ligand termini. This "misaligned" arrangement, shown in Figure 5b, begets a somewhat different PMF (see Figure 5a). The strong solvent layering we have described for passivated rods at low temperature is restricted to the neighborhood of sharp ligand-solvent interfaces. Near the ligand furrows, solvent structure is less pronounced. As a result, the approach of misaligned rods is not accompanied by strong overlap of intervening solvent layers. Oscillatory features of the "misaligned" PMF are therefore considerably less pronounced than in the aligned case. In addition, the deep minimum at contact is shallower and occurs at 3.4 nm (rather than 3.3 nm) separation due to weaker vdW attraction between the rods when the ordered regions on the opposing facets are misaligned. The mean force and the major force components for both the aligned and misaligned orientations are shown in Figure S1. Structure within the rods' ligand shells is only weakly sensitive to their relative alignment (see Figures S2-4).

Because nanorods have finite length, spontaneous symmetry-breaking in their ligand shells is not strictly possible. Over long enough time scales the free energy barrier for switching between furrow arrangements will be traversed, restoring the six-fold rotational symmetry of the disordered state. We have not computed the rate of this barrier-crossing, but observe that it does not occur on the ~10 ns time scale of our simulated trajectories. The aligned and misaligned states of approaching rods are thus meaningfully distinct in our simulations, but the true equilibrium state encompasses both. A PMF $\bar{\phi}_{MF}(r)$ that accounts for interconversion between these states can be simply computed by appropriately combining results $\phi_{MF}^{(A)}(r)$ and $\phi_{MF}^{(M)}(r)$ for the aligned and misaligned subensembles, respectively. Noting that the total partition function $\bar{Q}(r)$ for separation $r$ is a sum of aligned $Q^{(A)}(r)$ and misaligned $Q^{(M)}(r)$ contributions, and that the two states are equally probable at large separation, i.e., $Q^{(A)}(\infty) = Q^{(M)}(\infty)$, we obtain:

$$\bar{\phi}_{MF}(r) = -k_B T \, ln\left[\frac{e^{-\beta\phi_{MF}^{(A)}(r)} + e^{-\beta\phi_{MF}^{(M)}(r)}}{2}\right] \quad (5)$$

from the fundamental relationship $e^{-\beta\phi_{MF}(r)} = Q(r)/Q(\infty)$. This fully equilibrium result is also plotted in Figure 5a. It exhibits the deep contact minimum of the aligned state but a much reduced barrier between contact and the first solvent-separated minimum (where the misaligned state is typically more stable).

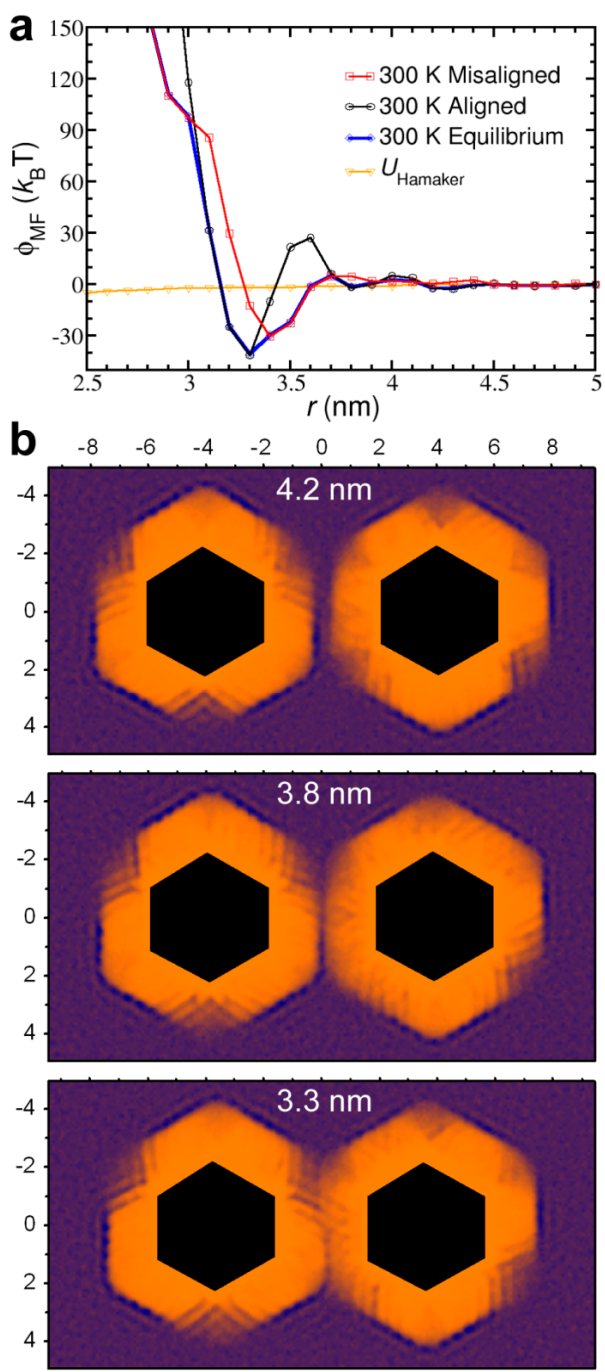

**Figure 5.** (a) Potentials of mean force $\phi_{MF}(r)$ for two parallel rods at 300 K in *n*-hexane with the ordered regions aligned and misaligned, where *r* is the distance between the opposing crystal facets. (b) Solvent densities for the misaligned case, averaged over time and along the 20 nm length of the CdS core, for several rod-rod separations. The scales are the same as for the solvent density in Figure 2.

So far we have only considered the PMF between CdS nanorods in a few select relative orientations, namely those in which the contact area between the rods is greatest. The actual affinity between the rods in solution will of course include contributions from many other relative orientations. Many of these we expect to be less attractive, even when the ligands are ordered. As a result, the overall attraction between rods at 300 K will be weaker than is suggested by the results we have presented. Nonetheless, attraction should be significant at low temperature, i.e., when the ligands are strongly ordered.

We have also calculated PMFs for a pair of passivated rods in vacuum at 300 K (Figure S6a), which underscore the key role of solvent in mediating interactions in solution. In the absence of solvent screening, the strength of dispersion interactions between nanoscale particles becomes apparent. These potent attractions considerably reduce the distance of minimum free energy, from ~3.3 nm in solution to ~2.4 nm in air, independent of the initial ligand configuration. This close approach requires significantly deforming the ligand shells (see Figure S6b). The primarily entropic cost of doing so is substantial (compare $F_{\text{Lig1-Lig2}}$ and $U_{\text{Lig1-Lig2}}$ in Figure S7), but is dramatically offset by dispersion, yielding a well depth of >600 $k_\text{B}T$ relative to the free energy of well-separated rods. We suggest that such contraction of the equilibrium rod-rod distance is likely the origin of cracking that is often observed in large dried assemblies[15]. In addition to a deep minimum, the PMF we have calculated for rods in vacuum features a less stable second minimum or pronounced shoulder (depending on temperature and alignment) near $r$ = 3.3nm, i.e. at the contact distance in solution when the ligands are ordered and aligned.

Our results reveal a mode of strong interaction among nanoparticles that is distinct from those typically considered, particularly in the context of self-assembly. This ligand- and solvent-mediated interaction thus adds to the repertoire of forces that could be tuned to promote pattern formation. It would be especially effective for designing structures that feature extensive face-to-face contact between faceted nanoparticles, as in exotic superlattices recently reported for octahedral Ag particles[54]. In that case depletion attraction due to polymers in solution afforded a modest bias for face-to-face contact, which in turn effected a dramatic change in superlattice structure. The interactions we have described in which solvent layering plays a significant role should favor parallel alignment of facets much more strongly, providing a versatile way to modify large-scale organization.

Due to the strong temperature dependence of the underlying ligand ordering transition, this alternative mode of face-to-face attraction can be switched on and off with precision and reversibility, and should make it possible to use thermal annealing to improve the quality of assemblies.

Interactions governed by spatial arrangements of passivating ligands should exhibit unique sensitivities to solution conditions and experimental protocols. Below we describe several such dependencies suggested by our computational results, discussing as well their implications for self-assembly of nanorods and other nanoparticles with large facets.

Changing the identity of the solvent can affect the mean force between faceted particles in several ways. Small flexible molecules like *n*-hexane order into discrete layers near smooth surfaces, resulting in short-ranged oscillatory solvation forces like the one that we observe when the ligands are highly ordered. Longer linear molecules and rigid spherical molecules will order more strongly[41,55], yielding solvent-mediated forces that are larger in magnitude and longer in range. By contrast, branched and irregular molecules, which do not form discrete layers, should produce a purely monotonic solvation force[41,55].

Associated changes in ligand solubility may have an even more dramatic effect. Reducing the solvent quality tends to cause aggregation of nanoparticles, a behavior sometimes exploited to drive the formation of ordered assemblies[6,56]. Our results suggest that the enhancement of effective particle-particle attraction can arise from multiple mechanisms. As traditionally discussed, a change in the balance of particle-solvent and solvent-solvent interactions mitigates the solvent's ability to screen bare attractions among nanoparticles or, in extreme cases, induces solvophobic forces that strongly drive aggregation. For example, low-resolution simulations of Au nanoparticles predict aggregation in poor solvents even in the absence of ligand ordering[57]. Our results suggest that changes in ligand solubility will additionally modify the compactness and order of the particles' passivation layers. As we have shown, the resulting changes in effective particle size and affinity that this causes can be a substantial additional effect.

The tendency of substrate-bound ligands to align is undoubtedly sensitive to the number bound per unit area. Many factors can influence the coverage of ligands on nanoparticles, including the identity of the crystalline core and its facet dimensions, the identity of the ligands (head-group, length, branching), and potentially even the preparation history (the solvent in which the particles are synthesized, the extent and time of washing, etc.). Generally, we expect reducing ligand coverage to raise the free energy of the ordered state and eventually to compromise its stability altogether. In work to be published separately,[49] we have explored the sensitivity of SAM ordering to reduced coverage for octadecyl ligands on the (100) facet of CdS. We find that the ordering transition persists (above 290 K) down to 75% coverage, broadening as ligands are removed. We therefore expect our results for nanorods to apply, at least qualitatively, to a broad range of experimental conditions.

The character of ligand ordering also depends sensitively on the size of facets to which they are bound. The most robust ordering, which upon cooling appears most sharply and at the highest temperature, is obtained for perfectly flat, macroscopic substrates. Our results indicate that edge effects introduced by the small size of nanocrystal facets broadens and lowers the transition temperature range in solution, much like reducing ligand coverage. This prediction is consistent with previous experimental[58–62] and numerical[34,35] studies of ligand ordering on Au, CdSe and silica particles (from 2-40 nm in diameter) in the absence of solvent. Overall, we expect the main result described in this paper, i.e. the transition from repulsive to attractive rods as the temperature is reduced below that of the ligand ordering transition, to have even greater relevance for particles with larger facets than the nanorods investigated here.

Another consequence of changing facet size (and thus particle curvature) is a change in the extent to which the ligand shells on different particles overlap when they come into contact. Previous work has shown that for small nanoparticles with substantial curvature the ligand shells tend to overlap significantly with one another in dry assemblies (e.g., 2-3 nm Au nanocrystals[30] or 5.5 nm Ag octahedra[63]); when the shell is disordered the ligands themselves interdigitate, while in the ordered state the overlap appears to occur mainly via the intercalation of aligned ligand bundles. This overlap causes the equilibrium particle separation to be fairly insensitive to the length of the passivating ligands[30,64]. In contrast, we do not observe significant overlap of the ligand shells on CdS nanorods at their equilibrium separation, irrespective of whether solvent is present or not. A similar lack of overlap was observed for organic acid ligands on 30 nm metal oxide nanoparticles[65]. These results indicate that it should be possible to tune the separation between particles with larger flatter facets by adjusting the ligand length. As some properties of nanoparticle films are highly sensitive to the core-core separation (e.g. electron transfer rates, which decrease exponentially with distance), this ability to tune the core-core separation will have important practical consequences for the design of nanoparticle-based materials.

Changing the ligand ordering should also change the surface energy of the nanorods, which will in turn affect how they assemble in the presence of fluid-fluid interfaces such as the solvent-air interface. It is well known that colloidal particles can be attracted to fluid-fluid interfaces, in particular when the balance of liquid-liquid and liquid-colloid surface free energies lowers the overall interfacial tension.[66] The strength of this attraction (or repulsion) depends crucially on the energies of the rod-fluid interfaces relative to the fluid-fluid interface. The nature of the rod-fluid interfaces will therefore be an important factor in determining how the rods assemble in the presence of such interfaces. Experimentally, CdSe/CdS nanorods assembled from toluene films on the surface of water have been observed to favor parallel alignment at ambient temperatures (~20 $^{\circ}$C)[14] and perpendicular alignment at elevated temperatures (45-60 $^{\circ}$C)[9]. While the precise reasons for this change in behavior are unknown, our results suggest that a change in the ligand ordering on the rods may be a significant factor.

To summarize, we have shown that octadecyl ligands on faceted nanorods in solution can undergo an ordering transition near room temperature, with the ligands forming large ordered domains that span the length of the nanorod. This ordering of the ligands in turn induces structuring of nearby *n*-hexane solvent and switches the rod-rod interaction from repulsive to strongly attractive. We have quantified this behavior through the potential of mean force acting between rods at a range of temperatures and for different relative ligand orientations. These results highlight the important and subtle roles of solvent in mediating interactions among passivated nanoparticles and surfaces, which would not be well described by standard implicit solvent models.

Numerous experimental and simulation studies have indicated that ligands on both isolated and clustered nanoparticles tend to locally order near room temperature in the absence of solvent, or in poor solvents.[28,31–37] To the best of our knowledge the current study is the first to demonstrate that ligands on nanoparticles can order near standard experimental temperatures even in good solvents, and also the first to describe how this order influences solvent structure and rod-rod interactions.

Our results are consistent with aspects of self-assembly phenomena involving passivated nanorods, for example the cracking that is commonly observed upon drying of nanorod films[15]. Our detailed predictions about ligand organization and rod interactions, however, await experimental verification. Directly observing these features in solution is extremely difficult, but surface-specific spectroscopic techniques (e.g., sum frequency generation[60,62]) offer promising routes to do so.

This work raises some new questions regarding ligand-mediated interactions: Does the ligand-order induced attraction that we have described here still occur when the ligands contain polar or charged end-groups? And how do the ordering temperature and surface forces vary as the nanoscale is approached, or in more practical terms, how small do the facets have to be before the attraction between ordered ligand layers is no longer significant? In work to be published separately, we have systematically characterized the effect of changing the facet dimensions and the ligand coverage on the ordering transition and the rod-rod interaction. Together, our results indicate that it should be possible to substantially tune the interaction between ligand-covered nanoparticles, and thus to affect their assembly into technologically useful structures, simply by varying the temperature.

**Acknowledgement.** This work was funded by the Helios Solar Energy Research Center, which is supported by the Director, Office of Science, Office of Basic Energy Sciences, Materials Sciences, and Engineering Division, of the U.S. Department of Energy under contract no. DE-AC02-05CH11231, and was supported by generous grants of computer time from the National Energy Research Scientific Computing Center, which is supported by the Office of Science of the U.S. Department of Energy under Contract No. DE-AC02-05CH11231. A.W. also acknowledges support from the Australian Research Council in the form of a Fellowship during the latter stages of this project.

**Supporting Information Available.** Details of methods used, and additional data on the interaction between rods in air and *n*-hexane. This material is available free of charge via the Internet at http://pubs.acs.org.

**Table of Contents Graphic**

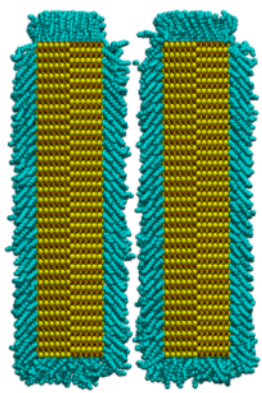 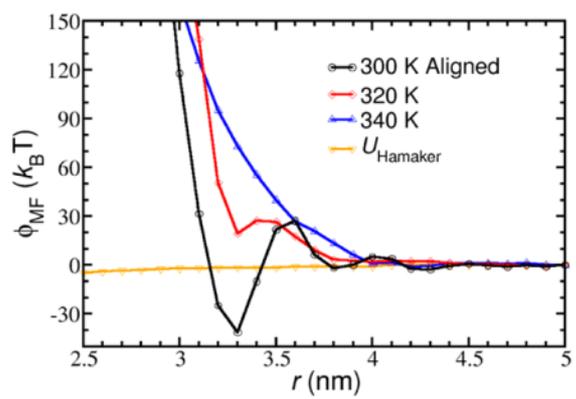